\documentclass[a4paper,fleqn]{cas-dc}

\usepackage[authoryear]{natbib}
\usepackage[abs]{overpic}
\usepackage{graphicx}
\usepackage{tikz}
\usepackage{amsmath}
\usepackage{xcolor}
\usepackage{tabularx}
\usepackage{pict2e}
\usepackage{multirow}
\usepackage[normalem]{ulem}

\def\tsc#1{\csdef{#1}{\textsc{\lowercase{#1}}\xspace}}
\tsc{WGM}
\tsc{QE}
\tsc{EP}
\tsc{PMS}
\tsc{BEC}
\tsc{DE}

\begin{document}
\let\WriteBookmarks\relax
\def\floatpagepagefraction{1}
\def\textpagefraction{.001}

\shorttitle{A simple trick for PIV/PTV}

\shortauthors{I. Tirelli et~al.}

\title [mode = title]{A simple trick to improve the accuracy of PIV/PTV data}                      



%
\author{Iacopo Tirelli}[
                        orcid=0000-0001-7623-1161]

\cormark[1]


\ead{iacopo.tirelli@uc3m.es}



\affiliation{organization={Department of Aerospace Engineering, Universidad Carlos III de Madrid},
            addressline={Avda. Universidad 30}, 
            city={Leganés},
            postcode={28911}, 
            state={Madrid},
            country={Spain}}

\credit{Methodology, Software, Validation, Formal analysis, Investigation, Data Curation, Writing - Original Draft, Visualisation}

\author{Andrea Ianiro}[orcid=0000-0001-7342-4814]
\credit{Conceptualisation, Methodology, Investigation, Writing - Original draft, Writing - Review \& Editing, Supervision}

\author{Stefano Discetti}[orcid=0000-0001-9025-1505]
\credit{Conceptualisation, Methodology, Software, Investigation, Resources, Writing - Original draft preparation, Writing - Review \& Editing, Supervision, Project Administration, Funding Acquisition}

\cortext[cor1]{Corresponding author}



\begin{abstract}
Particle Image Velocimetry (PIV) estimates velocities through correlations of particle images within interrogation windows, leading to a spatial modulation of the velocity field. Although in principle Particle Tracking Velocimetry (PTV) estimates locally a non-modulated particle displacement, to exploit the scattered data from PTV it is necessary to interpolate these data on a structured grid, which implies a spatial modulation effect that biases the resulting velocity field. This systematic error due to finite spatial resolution inevitably depends on the interrogation window size and on the interparticle spacing. It must be observed that all these operations (cross-correlation,direct interpolation or averaging in windows) induce modulation on both the mean and the fluctuating part. We introduce a simple trick to reduce this systematic error source of PIV/PTV measurements exploiting ensemble statistics. Ensemble Particle Tracking Velocimetry (EPTV) can be leveraged to obtain the high-resolution mean flow by merging the different instantaneous realisations. The mean flow can be estimated with EPTV, and the  fluctuating part can be measured from PIV/PTV. The high-resolution mean can then be superposed to the instantaneous fluctuating part to obtain velocity fields with lower systematic error. The methodology is validated against datasets with a progressively increasing level of complexity: two virtual experiments based on direct numerical simulations (DNS) of the wake of a fluidic pinball and a channel flow and the experimental data of a turbulent boundary layer. For all the cases both PTV and PIV are analysed.


\end{abstract}



\begin{keywords}
Particle Image Velocimetry \sep Particle Tracking Velocimetry \sep Turbulence statistics 
\end{keywords}

\maketitle

Cross-correlation-based Particle Image Velocimetry (PIV) is well known to suffer spatial resolution limits due to the averaging process within each interrogation window \cite{scarano2003theory,kahler2012resolution}. Recent efforts aimed at increasing the spatial resolution of flow field measurements are mostly focused on Particle Tracking (PT) approaches. The main argument in favour of PT is that the resolution limit is settled by the particle diameter \citep{kahler2012resolution}, and that the larger interparticle spacing in 3D reduces the risk of ambiguity in particle pairing if compared to its 2D counterpart. For 3D measurements, PT methods have demonstrated their superiority over cross-correlation-based methods  \citep{kahler2016main}. 

The main drawback is that PT provides velocity measurements only in the specific locations where the particles are available, leading to sparse sampling of the velocity field. In most applications (modal analysis, numerical differentiation, etc.) data are instead preferably represented in an Eulerian grid. This requires mapping the scattered data on a fixed grid. Several options can be considered, starting from a simple linear interpolation, or spatial weighted average \citep{agui1987performance}, up to more complex schemes enforcing physical constraints, such as the vortex-in-cell \citep{schneiders2014time}, the  ``FlowFit''  used in Shake-The-Box \citep{schanz2016shake} or constrained regression based on radial basis function \citep{sperotto2022meshless}. Independently from the complexity of the mapping scheme, this operation requires spatial information around each grid point on a region larger than the particle image diameter (i.e. the theoretical resolution limits).This introduces a systematic error in the form of modulation of the velocity fields which depends on the second-order spatial derivatives \citep{scarano2003theory} and on the region size. This form of error is analogous to the averaging effect observed in cross-correlation-based PIV, which depends on the interrogation window size.

Several solutions to reduce this form of systematic error have been introduced in the context of flow statistics (above all mean flow field and Reynolds stresses). Ensemble Particle Tracking Velocimetry (EPTV) exploits instantaneous realisations of spatially-scattered vectors (or trajectories) to build a dense cloud of vectors, from which local statistical distributions can be computed \citep{cowen1997hybrid,kahler2012resolution}. Local probability distribution functions (PDF) are obtained with bin averaging. Increasing the number of snapshots allows for decreasing the bin size progressively, thus improving the spatial resolution and reducing the modulation error. Sub-bin-size spatial resolution can be easily obtained by local polynomial fitting of the cloud of vectors \citep{aguera2016ensemble}, thus providing a further resolution enhancement and reducing spurious contributions to second-order statistics due to spatial gradients within the bin.

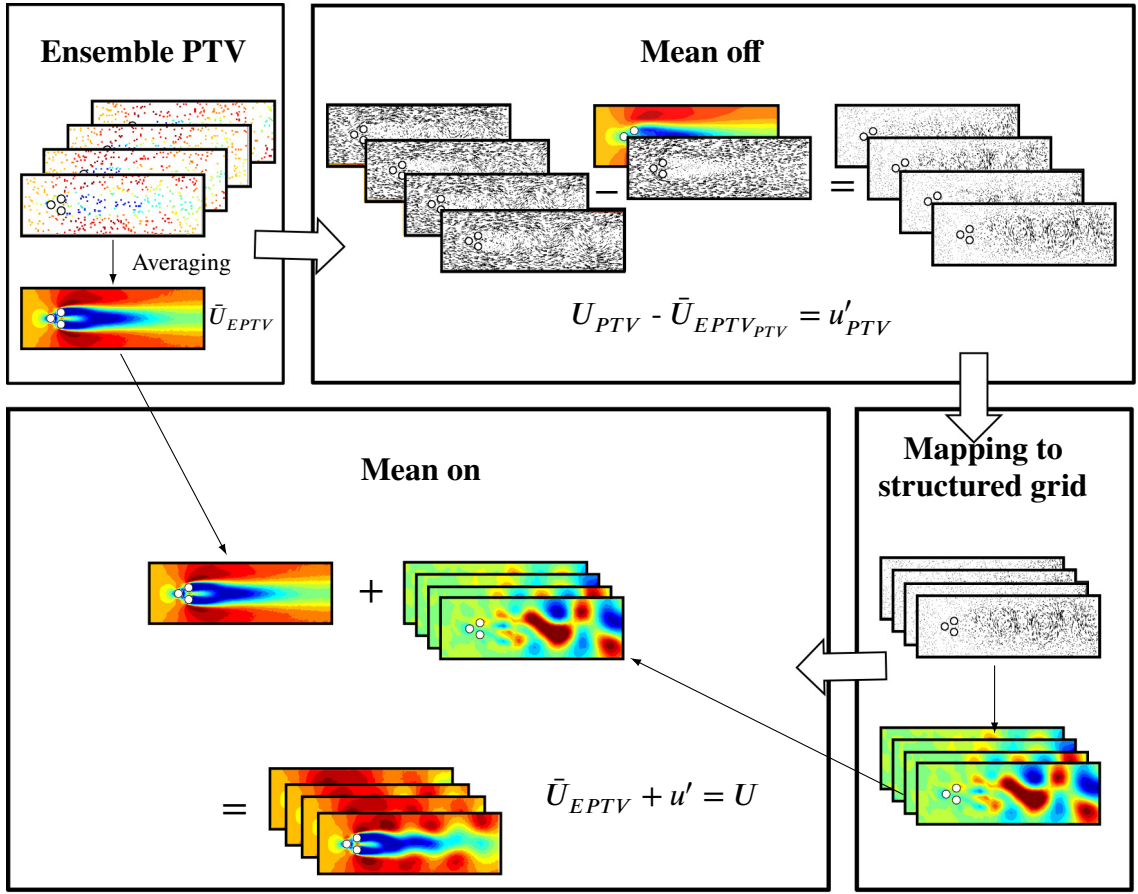
\begin{figure*}[t]
\centering 
\begin{overpic}[scale=1,unit=1mm]{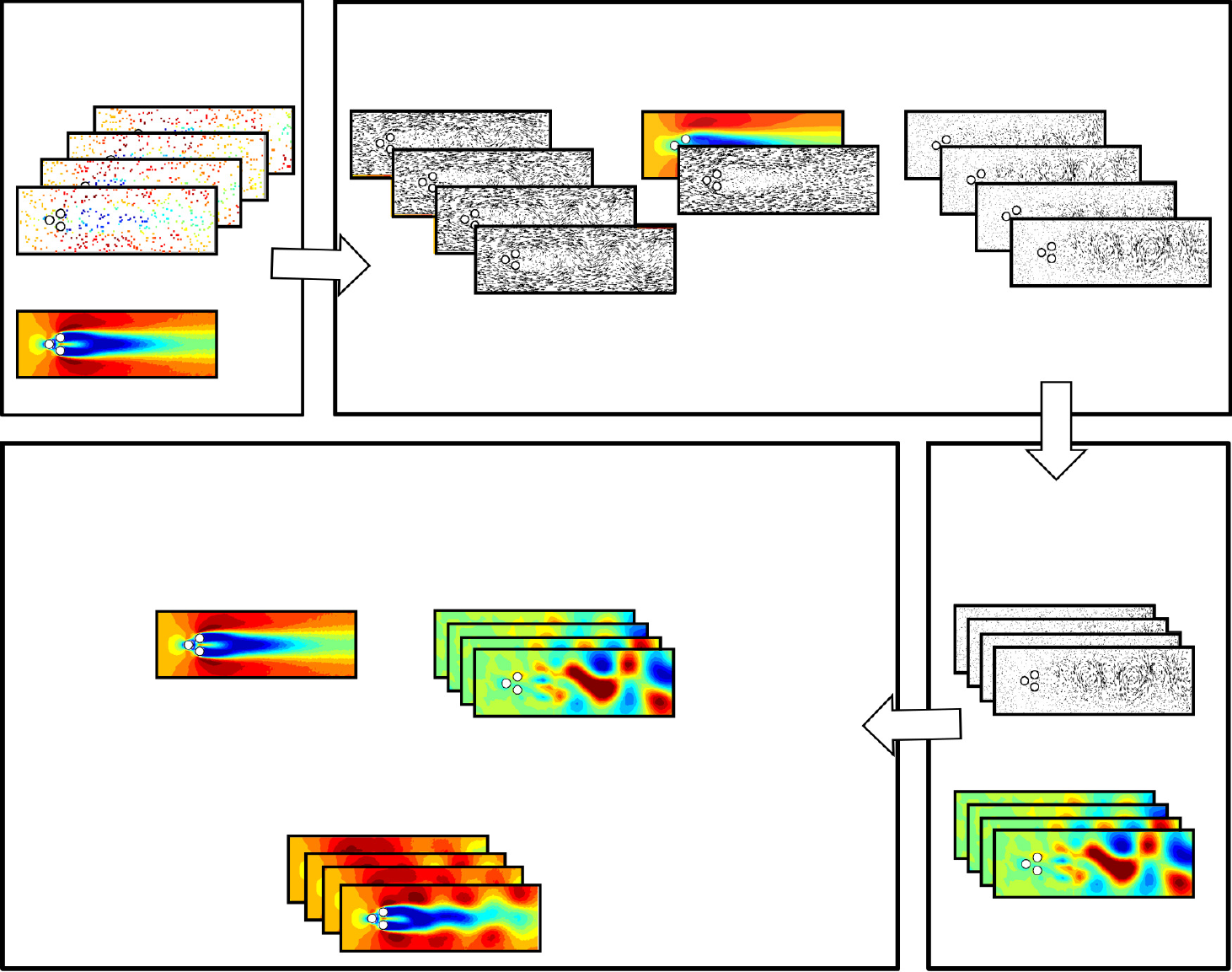}
\put(0,110){\parbox{35mm}{\centering \large\textnormal{\textbf{ Ensemble PTV }}}}
\put(14,86){\color{black}\vector(0,-10){6}}
\put(21,75){\parbox{20mm}{\centering \textnormal{$\bar{U}_{EPTV}$}}}
\put(13,82){\parbox{20mm}{\centering \textnormal{Averaging}}}
\put(41,110){\parbox{100mm}{\centering\large \textnormal{\textbf{ Mean off}}}}
\put(74,91){\parbox{10mm}{\centering {\Large{\textnormal{$-$}}}}}
\put(105,92){\parbox{10mm}{\centering {\Large{\textnormal{$=$}}}}}
\put(65,75){\parbox{60mm}{\centering {\large{\textnormal{\textbf{ $U_{PTV}$} - \textbf{$\bar{U}_{EPTV_{PTV}}$} = \textbf{$u^\prime_{PTV}$}}}}}}
\put(111,55){\parbox{35mm}{\centering\large \textnormal{\textbf{Mapping to structured grid}}}}
\put(130,30){\color{black}\vector(0,-10){9}}
\put(0,55){\parbox{109mm}{\centering \large\textnormal{\textbf{Mean on }}}}
\put(43,39){\parbox{10mm}{\centering {\Large{\textnormal{$+$}}}}}
\put(25,10){\parbox{10mm}{\centering {\Large{\textnormal{$=$}}}}}
\put(70,12){\parbox{30mm}{\centering {\large{\textnormal{$ \bar{U}_{EPTV} + u^\prime  =  U$ }}}}}
\put(118,13){\color{black}\vector(-80,40){36}}
\put(15,71){\color{black}\vector(19,-36){14}}

\end{overpic}
\caption{Flowchart of the proposed methodology. }
\label{Fig.flow}
\centering
\end{figure*}

While this process can be conducted on statistics, the equivalent bin size for the instantaneous field (which is dependent on the schemes used for mapping to the Eulerian grid) is somewhat related to the interparticle spacing. The bin size can thus not be easily reduced (or only to a certain extent, leveraging the recent methods proposed by \citealp{cortina2021sparse}, \citealp{guemes2022super} and \citealp{tirelli2023end} for snapshot PIV, or converting temporal into spatial resolution when time-resolved measurements are available, as in the works by  \citealp{lynch2013high,schneiders2016dense,schanz2016shake}). When mapping the instantaneous vectors onto a structured grid, or in the standard cross-correlation-based interrogation process, \textbf{the mean and fluctuating flow field undergo the same modulation}. Similarly, in cross-correlation-based PIV, the filtering effect of the interrogation process on windows of finite size applies equally on the mean and the fluctuating part of the flow. This is an unnecessary limitation, that can be easily removed.

The simple trick we propose is to compute the mean and the fluctuating flow fields separately. This can be done by computing the instantaneous flow fields with the traditional process, and then replacing the original mean flow field computed from the instantaneous snapshots with the one extracted by EPTV, whose systematic error is certainly smaller. 
The proposed solution is based on the hypothesis that the flow is statistically stationary.
Furthermore, it should be remarked that, in case of highly-periodic flows, a triple decomposition can be leveraged. The EPTV approach here opens to the possibility of correcting also the phase-average, further reducing the error.

To explain the advantage of the proposed procedure with an example, consider a one-dimensional sinusoidal shear displacement, superposed to a zero-time-mean fluctuating sinusoidal displacement:

\begin{equation}
\begin{split}
    U(x,t) = \bar{U}(x) + u^{\prime}(x,t) = \ldots\\ = A_{\bar{U}}\sin\left( \frac{2\pi x}{\lambda_{\bar{U}}}\right) + f(t)A_{u^{\prime}}\sin\left( \frac{2\pi x}{\lambda_{u^{\prime}}}\right),
\end{split}
\end{equation}

\noindent where $\bar{U}(x)$ is the mean flow field, $u^{\prime}(x,t)$ is the fluctuation, $A_{\bar{U}}$ and $A_{u^{\prime}}$ are the amplitude of mean and fluctuation, $\lambda_{\bar{U}}$ and $\lambda_{u^{\prime}}$ the respective wavelengths, and $f(t)$ is a function with zero temporal mean.

Assuming that the flow field is estimated with averaging on a bin of size $b$. The estimated instantaneous field would be equal to  $\tilde{U}(x,t)$:
\begin{equation}
\begin{split}
 \tilde{U}(x,t) = M_{\bar{U}}\bar{U}(x) +  M_{u^{\prime}}u^{\prime}(x,t)  \\
  =  M_{\bar{U}}A_{\bar{U}}\sin\left( \frac{2\pi x}{\lambda_{\bar{U}}}\right) +  f(t)M_{u^{\prime}}A_{u^{\prime}}\sin\left( \frac{2\pi x}{\lambda_{u^{\prime}}}\right).
\end{split}
\end{equation}

\noindent with $M_{\bar{U}},M_{u^{\prime}}$ being respectively the modulation of the mean and fluctuating flow field, computed for instance as in \cite{astarita2007analysis}.

The absolute error $\varepsilon$ is defined as the difference between $U(x,t)$ and $\tilde{U}(x,t)$:
\begin{equation}
\begin{split}
   \varepsilon = U(x) - \tilde{U}(x) = \left(1 - M_{\bar{U}}\right)A_{\bar{U}}\sin\left( \frac{2\pi x}{\lambda_{\bar{U}}}\right) + \ldots\\
   f(t)\left(1 - M_{u^{\prime}}\right)A_{u^{\prime}}\sin\left( \frac{2\pi x}{\lambda_{u^{\prime}}}\right) = \varepsilon_{\bar{U}} + \varepsilon_{u^{\prime}},
   \end{split}
   \label{eq.error}
\end{equation}

where we can distinguish the two contributes to the global error: $\varepsilon_{\bar{U}}$, from the mean component, and $\varepsilon_{u^{\prime}}$, from the fluctuating one.
Eq.~\ref{eq.error} clearly shows that if  $M_{\bar{U}}$ tends to 1 the error source on the mean field $\varepsilon_{\bar{U}}$ tends to zero. This can be approached easily in practical applications with EPTV, for instance.

\vspace{0.5cm}
Fig.~\ref{Fig.flow} summarises the methodology. 
The first step is computing the high-resolution mean flow field with EPTV. The bin size in this process can be set following:
\begin{equation}
    b = \sqrt{\frac{N_p}{N_tN_{ppp}}},
\end{equation}
where  $b$ is the bin size, $N_t$ is the number of snapshots available, $N_p$ is the number of particles in each bin needed for the convergence of the statistics (see \citealp{kahler2012resolution,aguera2016ensemble} for some guidelines on the selection of this parameter) while $N_{ppp}$ is the particle image density, expressed in particles per pixel. 
The output of this step is a high-resolution mean flow $\bar{U}_{EPTV}$.

In the second step, the instantaneous fluctuating velocity vectors are computed by subtracting the high-resolution mean flow field obtained in the previous step. In the third step, the fluctuating velocity vectors are then mapped on the structured grid in case of PT approaches. This step can be skipped for standard cross-correlation-based PIV. Without loss of  generality, we consider this process being equivalent to a spatial weighted average over a kernel containing a reasonable number of vectors (at least $7-10$, following the standard rules of thumb of PIV). In any case, it can be safely assumed that the modulation effects will be significantly larger on the instantaneous fields if compared to the EPTV process. Carrying out the mapping on the structured grid only for the fluctuating velocity component allows to have modulation effects only on the fluctuating velocity component.

In the last step, complete instantaneous velocity fields are obtained by the superposition of the high-resolution mean flow $\bar{U}_{EPTV}$ to the fluctuating velocity fields. Recalling that $\varepsilon_{\bar{U}}$ is negligible thanks to the EPTV process, this procedure leads to a smaller $\varepsilon$ on the instantaneous velocity fields.

 \begin{figure*}[t]
\centering 
\begin{overpic}[scale=1,unit=1mm]{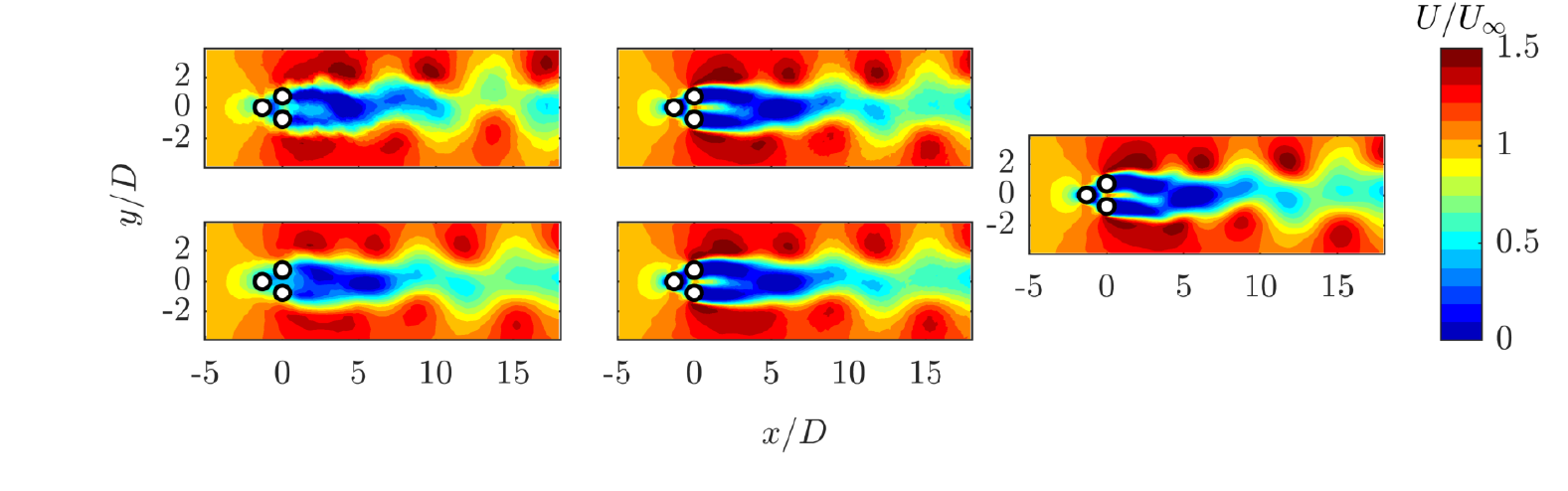}
\put(20,50){\parbox{35mm}{\centering \textnormal{Standard process}}}
\put(60,50){\parbox{40mm}{\centering \textnormal{After mean correction}}}
\put(100,50){\parbox{40mm}{\centering \textnormal{Reference}}}
\put(-10,20){\parbox{20mm}{\flushleft \textnormal{PIV IW = $32$}}}
\put(-10,40){\parbox{20mm}{\flushleft\textnormal{PTV b = $32$}}}
\put(-10,48){\parbox{60mm}{ \flushleft \textnormal{Testcase 1: fluidic pinball}}}
\put(0,0){\color{black}\linethickness{0.3mm}
\polygon(21,27.5)(35,27.5)(35,15.5)(21,15.5)}
\put(0,0){\color{black}\linethickness{0.3mm}
\polygon(21,45)(35,45)(35,33)(21,33)}
\end{overpic}

\begin{overpic}[scale=1,unit=1mm]{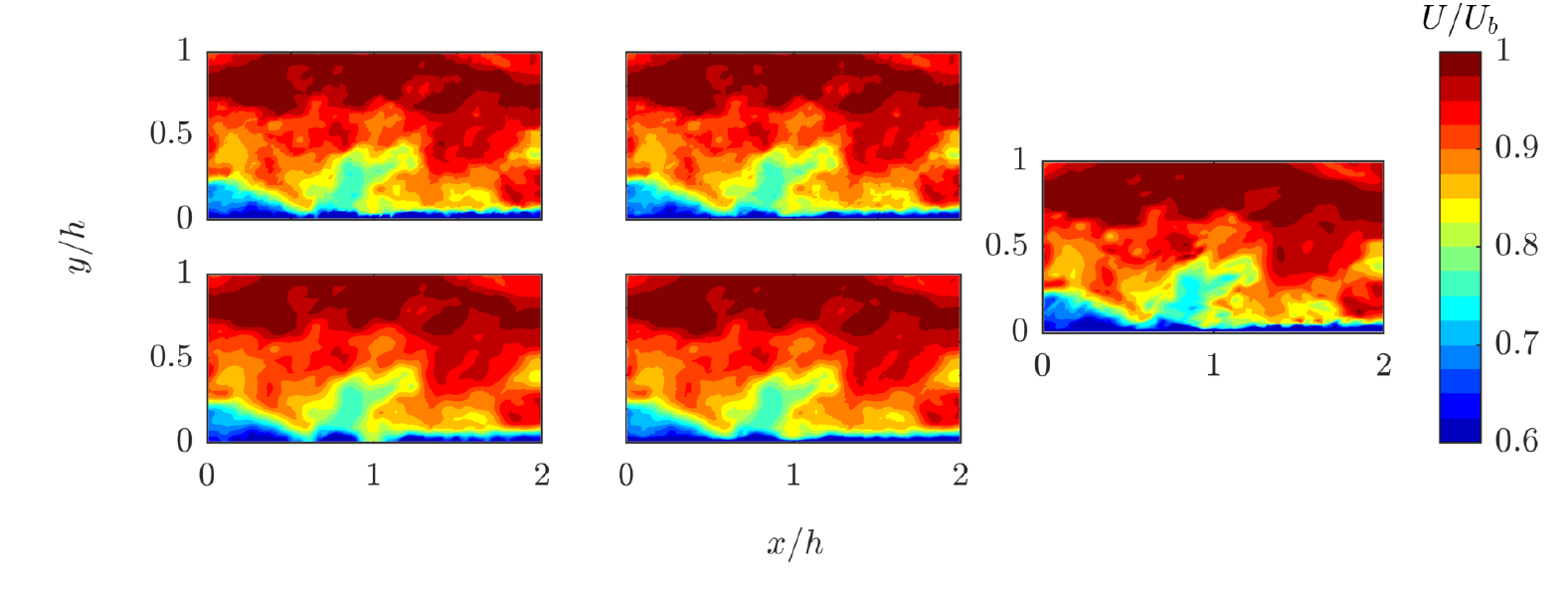}
\put(-10,25){\parbox{20mm}{\flushleft \textnormal{PIV IW = $32$}}}
\put(-10,48){\parbox{20mm}{\flushleft \textnormal{PTV b = $32$}}}
\put(-10,60){\parbox{60mm}{ \flushleft \textnormal{Testcase 2: turbulent channel flow}}}
\put(0,0){\color{black}\linethickness{0.3mm}
\polygon(21,15)(55,15)(55,17)(21,17)}
\put(0,0){\color{black}\linethickness{0.3mm}
\polygon(21,38)(55,38)(55,40)(21,40)}

\end{overpic}

\begin{overpic}[scale=1,unit=1mm]{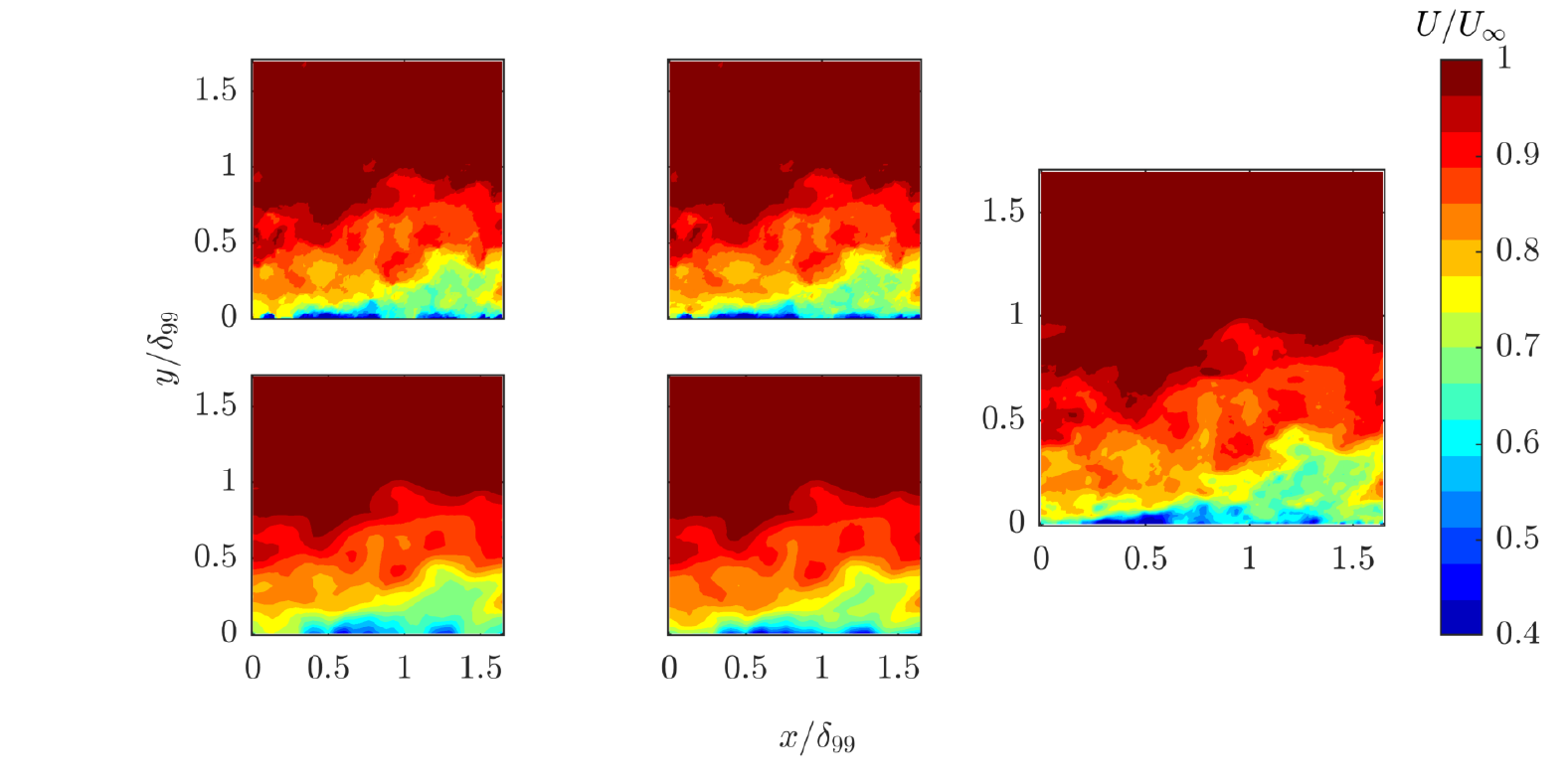}
\put(-10,28){\parbox{20mm}{\flushleft \textnormal{PIV IW = $128$}}}
\put(-10,60){\parbox{20mm}{\flushleft \textnormal{PTV b = $128$}}}
\put(-10,80){\parbox{60mm}{ \flushleft \textnormal{Testcase 3: turbulent boundary layer}}}
\put(0,0){\color{black}\linethickness{0.3mm}
\polygon(25.5,15.5)(51.5,15.5)(51.5,17.5)(25.5,17.5)}
\put(0,0){\color{black}\linethickness{0.3mm}
\polygon(25.5,47.5)(51.5,47.5)(51.5,50)(25.5,50)}
\end{overpic}
\centering
\caption{ Panel  with different examples of applications of the method. The top rows picture the PTV cases, while the bottom one the PIV cases, in both of them instantaneous streamwise velocity field contours are displayed.
 In the left column, there are the initial flow fields, before the application of the methodology, while the central column represents the flow fields after substituting the mean. The last column reports the references for each case: DNS for the virtual test cases and PIV with IW =  $32$ pixels for  the turbulent boundary layer. The rectangles pictured in the first column represent the smaller regions where the local error is computed: $x/D \in [-5D,4.5D]$ for the fluidic pinball, $y/h \in [0, 0.17]$ for the turbulent channel flow and  $y/\delta_{99} \in [0, 0.13]$ for TBL. }
\label{Fig.panel}
\end{figure*}
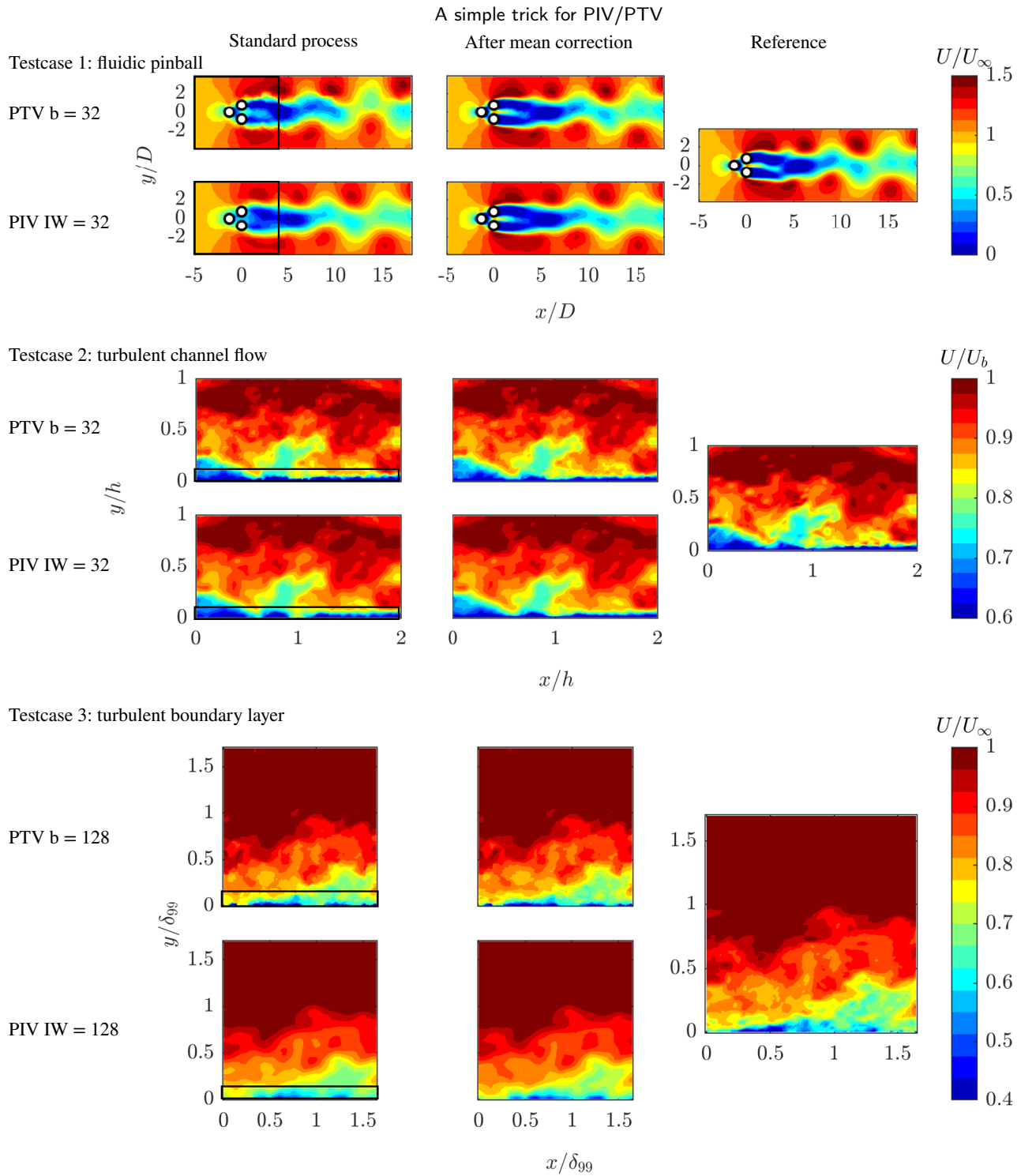

\vspace{0.5cm}
The methodology is validated for PIV and PTV cases with different datasets with an increasing level of complexity: the flow around the fluidic pinball \citep{deng2020low}, a turbulent channel flow \citep{li2008public} and the experimental dataset of a turbulent boundary layer \citep{guemes2019experimental}. The virtual PTV datasets of the pinball and the channel flow test cases are obtained imposing that the exact position of the particles is known. For the corresponding cross-correlation analysis, instead, synthetic images are  by locating on these positions Gaussian-shaped particles with a diameter of $2$ pixels and maximum intensity of $100$ counts. The synthetic images have been processed with a standard multi-step window deformation PIV analysis \citep{scarano2000advances}. Regarding the experimental data, they are collected in the facility of the Aerospace Engineering Group at Universidad Carlos III de Madrid. The velocity vectors are extracted with a super-resolution PTV approach \citep{keane1995super}. The datasets tested in this work are those employed by \cite{tirelli2023end}; the reader is referred to it for further details.

 \begin{table*}[t]
\centering
\begin{tabular}{ ||l|c|c|c|c|c|c|c|c|| }
 \hline
 \multicolumn{5}{||c|}{$\langle\delta_{RMS}\rangle$}& \multicolumn{4}{|c||}{$\langle\delta_{RMS_{local}}\rangle$} \\
  \hline
  \multicolumn{1}{||c|}{} & \multicolumn{2}{|c|}{PTV}& \multicolumn{2}{|c|}{PIV} & \multicolumn{2}{|c|}{PTV}& \multicolumn{2}{|c||}{PIV}\\ 
   \hline
  Testcase   & Standard    &Corrected & Standard    &Corrected  & Standard    &Corrected & Standard    &Corrected\\
 \hline

 Fluidic pinball   & 0.0628   &0.0389 & 0.0830   &0.0528 & 0.0781  & 0.0267 & 0.1042 & 0.0444\\
 Turbulent Channel flow & 0.0220  & 0.0198 & 0.0248    &0.0222 &0.0542 &0.0418 &0.0638 &0.0491 \\
TBL & 0.0219 &0.0213 & 0.0218    & 0.0202 & 0.0733  & 0.0672 & 0,0663 &0.0573\\
 \hline
\end{tabular}
 \caption{Spatial average of the root mean square error $\langle\delta_{RMS}\rangle$ evaluated for all the testcases and both the techniques for the standard process and after applying the mean flow correction. The local root mean square error  $\langle\delta_{RMS_{local}}\rangle$ refers to evaluation over small regions localised by large errors, namely $x/D \in [-5D,4.5D]$ for the fluidic pinball, $y/h \in [0, 0.17]$ for the turbulent channel flow and  $y/\delta_{99} \in [0, 0.13]$ for the turbulent boundary layer. In all the testcases and for both the approaches (PIV or PTV) the first column shows the error (global or local) obtained without performing the substitution, while the other one the updated error performing the methodology. The number of snapshots exploited for EPTV are: $4737$ (fluidic pinball), $11856$ (turbulent channel flow) and $20000$ (TBL). }
\label{tab.err}
\end{table*}

The metric used to assess the improvements given by the current methodology  is the normalised root mean square error $\delta_{RMS}$, evaluated as:

\begin{equation}\label{eq.rms}
   \delta_{RMS} = \frac{\sqrt{\frac{\sum\limits_{i = 1}^{N_t}(U_i-U_{Ref_i})^2 + (V_i-V_{Ref_i})^2}{N_t}} }{U_\infty},
\end{equation}
 where $U$ and $V$ are the PIV/PTV velocity vectors, $U_{Ref}$ and $V_{Ref}$ are the corresponding reference vectors (Direct numerical simulation (DNS) for the synthetic cases and PIV with interrogation window IW = $32$ pixels for the experimental one), $N_t$ is the number of snapshots and $U_\infty$ is the freestream velocity.

 The panel in Fig.~\ref{Fig.panel} summarises all the different scenarios explored during the validation. Dimensionless coordinates are obtained using the diameter $D$ of the cylinders, the half-channel-heights $h$ and the boundary layer thickness $\delta_{99}$. The left column of the panel shows the contours of the streamwise velocity fields with the standard process, while the central one illustrates the same field after the mean-flow correction. The reference is reported in the last column. Local analysis covers the area outlined with black boundaries in the first column. In all the cases the first  row collects the results of the PTV analysis while the second those of standard cross-correlation-based PIV. 
 
 For the fluidic pinball, the panel shows a comparison with the mapping of the PTV vectors on the regular grid using a top-hat averaging over a bin size of $32$ pixels. The PIV process has the same final interrogation window size. The total amount of snapshots included is $4737$ and the freestream velocity $U_\infty$ is equal to $1$ pixel. Due to the moving average process, the PIV appears to cancel out the smallest scales in the first column, especially in the restricted area for local analysis. Table~\ref{tab.err} reports the root-mean-square errors for all the applications. PTV and PIV have respectively an initial error of $0.0628$ (PTV) and $0.0830$ (PIV) on the entire field while in the restricted region that spans on the entire crosswise direction and $x/D \in [-5D,4.5D]$ the error increases to $0.0781$ and $0.1042$ respectively. The second column illustrates the benefits of correcting the mean flow using EPTV with a bin size of $3$ pixels. As proof of this, the local rms is drastically reduced up to $0.0267$ and $0.0444$ respectively, while on the entire dataset it is $0.0389$ and $0.0528$.

The turbulent channel flow confirms this trend, but this time the improvement is less evident. In this test case a total of $11856$ snapshots has been employed and the parameter used for dimensionless measurements is the bulk velocity $U_b = 7.5 $ pixels. The effect of the high-resolution mean is to recover the smallest scales that are cancelled out, in particular near the wall.  This finds confirmation focusing in the region $y/h \in [0, 0.17]$ (Table~\ref{tab.err}). The PIV error is higher than the PTV one in this specific region because it is dominated by the smallest scales. 

The last test case is based on experimental PIV data from a turbulent boundary layer. As in \cite{tirelli2023end}, the number of vectors per snapshot has been artificially reduced by a factor of $10$. This allows generating a robust ground truth with PIV processing on a $32\times32$ pixels window using the original images, while the process analysed here is performed on downsampled images with (on average) $10$ particles on a $128\times128$ pixels window.The freestream displacement in this last testcase is 12.8 pixels. It should be remarked that, in order to have a fair comparison with the reference fields, the EPTV process for this testcase is limited to a bin equal to the interrogation window of the PIV used as reference, i.e. $32$ pixels. This process is necessary to avoid a misleading resolution difference between the mean field from the reference PIV data and that from the EPTV, which would be reflected in a spurious error component. On the other hand, it must be noted that, with the current dataset, it would be possible to obtain a more accurate mean field with EPTV by pushing the bin size to significantly smaller values.
 
In all these different test cases, the methodology shows a significant improvement of the spatial resolution, thanks to the minimisation of the systematic error associated with the mean flow component $\varepsilon_{\bar{U}}$.

It is evident that the larger is the size difference between window used in the interrogation for the instantaneous fields and bin used for the EPTV, the larger is the improvement to be expected. If the window for the instantaneous field estimation would be sufficiently small to have a negligible error on the mean field, the improvement would be negligible.

The range of applications of this correction is wide, and it can be easily incorporated in all the processing procedures usually adopted in PIV/PTV, providing also additional advantages. For instance, techniques such as the KNN-PTV \citep{tirelli2023end} depend on the low-resolution first guess, which can be substantially improved with this procedure.  In the test cases proposed here, the error of KNN-PTV decreases from 0.0299 up to 0.0288 for the fluidic pinball and from 0.0207 to 0.0196 for the channel flow when including the mean-flow correction in the process.

 \vspace{0.5cm}
 In conclusion, this manuscript presents as a simple correction for PIV measurements to minimise the systematic error due to finite spatial resolution. The solution is based on a mean-flow shifting using the high-resolution mean computed with ensemble methods. The systematic errors due to large window size in instantaneous field measurements will thus impact only the fluctuating part of the velocity field.  
 The effectiveness has been tested and validated in different conditions, with robust improvement in all tested cases. Provided that the bin size for the computation of EPTV is not pushed to a level that affects convergence, an improvement is expected to be observed in all testing conditions. This methodology is very easy to implement and computationally inexpensive. The computational cost added by the application of the proposed mean-flow correction is negligible (assuming that EPTV has been already carried out to obtain high-resolution statistics). The interested reader can refer to the appendix of \citet{tirelli2023end} for an estimation of the computational cost of the typical steps involved in the process.
 
 The proposed method provides more accurate results, which can be better exploited, for instance, for numerical code benchmarking, model development, or data assimilation in simulations.
 The resolution of each velocity measurement is improved with minimal effort, thus we recommend including this correction in all PIV/PTV algorithms.

\vspace{0.1cm}

\printcredits

\section*{Declaration of completing interest}
The authors declare that they have no known competing financial interests or personal relationships that could have appeared to
influence the work reported in this paper.

\section*{Data availability}
All datasets used in this work are openly available in Zenodo, accessible at \url{https://doi.org/10.5281/zenodo.6922577}. 

\section*{Acknowledgment}
This project has received funding from the European Research
Council (ERC) under the European Union’s Horizon 2020 research and innovation program (grant agreement No 949085). Funding for APC: Universidad Carlos III de Madrid (Read \& Publish Agreement CRUE-CSIC 2022). 

\bibliographystyle{cas-model2-names}
\bibliography{cas-refs}

\end{document}